\documentstyle[aps,preprint]{revtex}

\begin{document}

\author{
Daniele Passerone$^{a,b}$\cite{corrauth}, Furio Ercolessi$^{a,b}$\cite{aff},
 Franck Celestini$^c$\cite{aff2}, and Erio Tosatti$^{a,b,d}$\cite{aff3} \\
{\it a) Istituto Nazionale per la Fisica della Materia (INFM)}\\
{\it b) International School for Advanced Studies, Trieste, Italy}\\
{\it c)  Laboratoire MATOP associ\'e au CNRS, Universit\'e
d'Aix-Marseille III,
Marseille, France} 
{\it d) The Abdus Salam International Centre for Theoretical Physics,}\\
{\it Trieste, Italy}\\
}
\title
{Realistic simulations of Au(100): Grand Canonical Monte Carlo 
and Molecular Dynamics}
\maketitle
\begin{abstract}

The large surface density changes associated with the (100) noble metals 
surface hex-reconstruction suggest the use of
non-particle conserving simulation methods. We present an example
of a surface Grand Canonical Monte Carlo applied to the transformation
of a square non reconstructed surface to the hexagonally covered 
low temperature stable Au(100). On the other hand, classical 
Molecular Dynamics allows
to investigate microscopic details of the reconstruction dynamics, and we
show, as an example, retraction of a step and its interplay 
with the surface reconstruction/deconstruction mechanism. 
\end  {abstract}

\newpage

The (100) surface of noble metals (Au, Pt, Ir) shows 
a quasi-hexagonal reconstruction, which has been deeply investigated 
both experimentally \cite {olderexpAu,expAu} and theoretically \cite{theoAu}.
 The reconstruction is characterized by an approximate 
 (in Au and Pt, where a slight incommensurability is observed) or 
 exact (in Ir)  $(1 \times 5)$  unit cell, arising from the interplay
 of a slightly contracted and distorted triangular surface layer
 with the (square) second layer.  This result can be interpreted in
 terms of a simple model (the {\em glue model}) incorporating the
 tendency of surface atoms to increase their coordinations \cite{theoAu}.
 As the result of a strong packing tendency, the density of the top layer
 at low temperature is about 24\% larger than that of a regular (100)
 layer, and it tends to increase with temperature \cite{sandy}.
 Moreover, the angular orientation of the top layer undergoes 
 phase transitions well characterised by X-ray measurements \cite{xrayang}.

 At a temperature of $0.81 T_m$, a deconstruction transition occurs, 
 and the long-range hexagonal order parameter vanishes. 
 A surprisingly exact coincidence of  (rescaled) transition temperatures 
 for the three noble metals is observed.
 In spite of this large amount of theoretical and experimental work, 
 a detailed insight into the consequences of the strong density difference 
 between a square, unreconstructed $(100)$ layer and the reconstructed 
 triangular surface is still lacking, particularly concerning the interplay 
 between defects such as steps, and the reconstruction/deconstruction 
 mechanism. To this end, Molecular Dynamics (MD)
 is a valid tool, especially for noble metals such as gold, where 
 glue potentials have revealed to be predictive.
 Classical MD makes it possible to simulate a large number of atoms for 
 a time scale of several nanoseconds, allowing for instance to follow
 the movement of a step on a surface.
 On the other side, a deep theoretical understanding of the deconstruction 
 transition at high temperatures has not been achieved until now, 
 and MD simulations have a difficulty in reproducing this aspect 
 due to the atom number conservation constraint: strong density changes 
 at the surface are involved, and crater/islands appear as a consequence 
 of these changes and of the impossibility to create/delete atoms as needed.
A more natural method in this case should be Grand Canonical Monte Carlo
(GCMC).
Recently, we have successfully applied this method to the preroughening 
of rare gas crystal surfaces \cite{Prcelest} and to rare gas adsorption 
on an attractive substrate \cite{Prads}.
 In the case of metals the low vapor pressure and equilibration problems 
 hinder in some way the success of the method. Preliminary results 
 are however encouraging.
 
In this paper we present two examples of simulation on the Au(100) surface.
The first is a calculation which reveals the possibilities of GCMC 
even in the case of metals: starting from a square surface, 
the system spontaneously formed a hexagonally reconstructed overlayer; 
the second example is a high-T MD simulation of a step on this surface, 
and its interplay with the reconstruction.
In both studies we have adopted a slab 
geometry with Periodic Boundary Conditions along the $x-y$
directions. The potential we used is the {\em glue} potential \cite{glue}
which consists of a two-body term and of a density dependent term,
mimicking the valence electrons effects in metals.

First, we will report on our GCMC study. 
Our Monte Carlo procedure involves small displacement moves (m),
creations (c), and destructions (d) with relative probabilities
$\alpha^{(m)}=1-2\alpha$ and $\alpha^{(c)}=\alpha^{(d)}=\alpha$,
with $\alpha\simeq 0.25$.
Small moves apply to all particles, whereas creation/destruction
is restricted to a fixed surface region, about four layers wide.               
We have simulated two different systems at low temperature: 
a non-reconstructed surface (A) and a reconstructed quasi-hexagonal 
surface (B) 
In both cases the slab had 225 atoms per layer, and was 16 layer thick. 
We found the following results:

\begin{enumerate}
\item  upon adsorption of atoms on the square top layer [system (A)], 
the adatoms are included and the equilibrium state of the surface is, 
correctly, the $(5\times 1)$ reconstructed quasi-hexagonal state; 

\item  upon adsorption of atoms on the reconstructed layer [system (B)], 
another reconstructed layer forms, and the former first layer, now covered, 
deconstructs immediately into a nearly regular, square $(100)$ layer.
\end{enumerate}

The surface density $\rho _{\left( s\right) }$ for system (A) 
continuously changes from the bulk value to the hexagonal value. 
The chemical potential is $\mu =-3.92 {\rm eV}$ and the averaged density
$\langle\rho _{\left( s\right) }\rangle$ levels up at a nearly 
correct value of $1.30$. All simulations are performed at $T=800\,K.$ 
Figure 1 shows the initial unreconstructed state (A), two intermediate states with hexagonal zones which begin to form, and the equilibrium final state, triangularly ordered.

For system (B) the evolution of the number of atoms of the ad-layer and the population of the first two adsorbed layers is shown in Figure 2. The density of the first layer decreases nearly to the bulk value when the ad-layer has completed its growth, and completes its square ordering when another adlayer has adsorbed on it.

These results show that the GCMC simulation method is efficient also 
on the microscopic scale: structural phase transitions involving 
strong density changes and rearrangement of atoms are correctly driven 
by the grand canonical Markovian chain we used in order to sample 
this particular statistical ensemble.
Not only the correct density is achieved, but also an almost perfect 
short-range order is obtained both in the reconstructed top layer 
and in the square bulk layers. This fact is remarkable, 
if one takes into account the fact that an off-lattice model was used 
and no constraints on the lattice structure were given.

Let us now present the other study concerning Molecular Dynamics simulation 
of a step on a flat Au(100) surface. We adopted the same glue potential 
as before in the slab geometry (16 layer slab, with a maximum of 2500 atoms 
per layer), numerically integrating Newton's equations for simulation 
times of the order of nanoseconds.

Upon increase of temperature, the lateral density of a flat reconstructed 
Au(100) surface shows a tendency to increase (enhancement of lateral coordination is compensated by an anomalous outwards relaxation of the first layer). We find in our simulations 
\cite {noiss} an increase of the T $=0$ lateral density of 1.24 
with respect to the bulk, to 1.35 for T $=1000K$. This behavior is 
in close agreement with experiment for both Au and Pt \cite{expAu}.

When a step is added to the flat surface, it becomes a free source of atoms 
for a layer which needs to increase its density. Heating can therefore 
trigger step retraction on a hex-reconstructed surface.
We have studied the mechanism of this retraction in Au(100) in a 
`one step geometry'. Usual periodic boundary condition in the $x-y$ plane 
of the slab are not suitable to simulate a surface with a single step:
only an even number of steps allows matching at the boundaries,
due to the $ABABAB...$ stacking of $(100)$ layers in a fcc crystal. 
In order to allow modelling a single step, we have adopted special 
boundary conditions: if an atom of a layer $A$ crosses the slab boundary 
which is parallel to the step, it appears on the
other side displaced vertically by one layer, with the proper horizontal 
mismatch between layer $A$ and layer $B$. In this way, a single step 
separates an $A$ terrace from a $B$ terrace. 

When the step retracts, it uncovers by doing so a portion of terrace 
which in turn must also become reconstructed. This must occur most naturally 
by incorporation into the lower terrace of atoms formerly belonging 
to the upper retracting terrace, whose retraction must in turn be accelerated 
by this loss.
Conversely, when a step advances, it covers a portion of terrace which 
must at the same time deconstruct. This must liberate excess atoms, 
which must in their turn be incorporated in the advancing upper terrace, 
causing it to advance even more. While this ``positive feedback'' is obvious, 
the mechanisms of its actual occurrence have not been described before, 
and their consequences seem worth studying in some detail.
The starting point of our simulations was an equilibrated step at T=$800 K$ 
(Figure 3) then suddenly brought to a higher temperature of $950 K$. 
The equilibrium lateral density of the upper surface suddenly jumps 
upwards by 5\%, whence the step begins to retract. We show in Figure 4 
some stages of the simulation. To clarify the process we have marked 
in yellow the top layer atoms, labelled the atoms which participate 
in the process, marked with a number the initially top-layer atoms, 
and with a letter 
the atoms initially belonging to the second layer.  
After 70 ps (Figure 5), retraction has occurred and the uncovered zone 
has become reconstructed: atoms which were formerly on the step edge 
(atoms 1-7 and 11-15), have been incorporated into the lower terrace 
and form with the atoms of the formerly quadratic substrate a 
hexagonal pattern. The positive feedback which we had anticipated does exist, 
and appears also to play a dynamical role in rendering 
the retracting step particularly wiggly.

In conclusion, we have presented two examples of simulation on the Au$(100)$ 
surface. 
Concerning the GCMC exemplification, we  must give, of course, 
credit to the glue potential we have used \cite{glue}, but we are confident
that the interplay between efficient phenomenological potentials 
and particle-non-conserving algorithms such as the one just described 
will lead to interesting predictions in the field of structural transitions 
at surfaces, in particular because no assumptions on the final structure
need to be made in advance.
Summarizing, this simple test indicates the feasibility of a surface GCMC 
simulation, applied in particular to metals modeled by classical many-body 
(glue) forces.

The single step simulation shows an example of non-standard 
Periodic Boundary Conditions, and the power of classical MD for the 
investigation of microscopic details of physical phenomena.
Single atoms can  be `followed' during the simulation; in this case, 
an interesting interplay between step movement, density change and 
reconstruction has been exploited.

We thank Francesco Di Tolla for discussions. Work at SISSA by D. P. is 
directly supported by MURST-COFIN 97.

\section{Figure captions}

Fig. 1:(a) The initial state of Au(100): the top layer is perfectly
square and unreconstructed; (b) the Monte Carlo creation of new particles 
starts modifying the structure; (c) the correct lateral density is achieved; 
(d) quasihexagonal order (equilibrium state) is obtained at surface 
after the Monte Carlo simulation. Color of atoms reflects their height, 
atoms in (d) are brighter due to vertical expansion (about $20\%$) 
connected to the reconstruction of the first layer.

Fig. 2: Evolution of the particle occupancy of the initially first layer 
and of two layers growing on it (system (B)). Layer densities are normalised 
to bulk $(100)$ lateral density. The first layer decreases its density 
and becomes square with defects, and eventually perfectly square 
when two complete layers are adsorbed on it. The deconstruction of 
the underlying layer increases the growth rate of an adlayer.

Fig. 3 (color): Top view of a detail of the simulated slab. 
Equilibrated step at $800 K$, suddenly brought to 
a higher temperature of $950 K$. Yellow atoms are top layer atoms, 
blue atoms belong to the second layer. 

Fig. 4 (color): Intermediate snapshots of the simulation 
(frames are separated by 1.4 ps). The step has retracted and a whole line 
of atoms (1-7) passed from the step rim to the second layer. 
Atoms 13, 14, 5, 6 and 7 are part of the reconstruction of 
the uncovered zone of the second layer. 

Fig. 5 (color): The final situation, with evident shrinking of the step 
(70 ps of simulation have been completed). The uncovered zone 
has become reconstructed, and atoms $1-7$, formerly at the step edge, 
as well as $11-15$, formerly second
line, have been incorporated into the lower terrace. 
Also note the new large wiggliness of the retracted step.

\end {document}
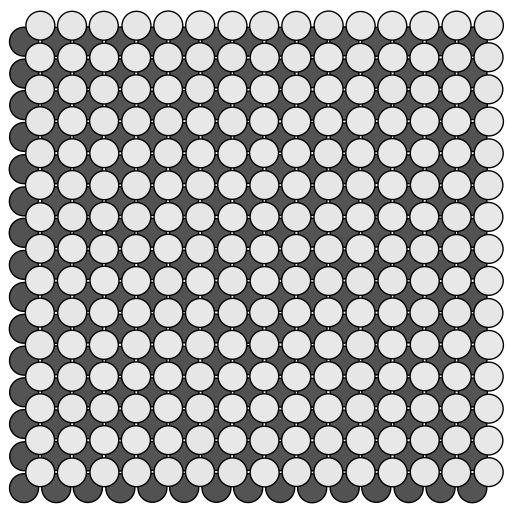
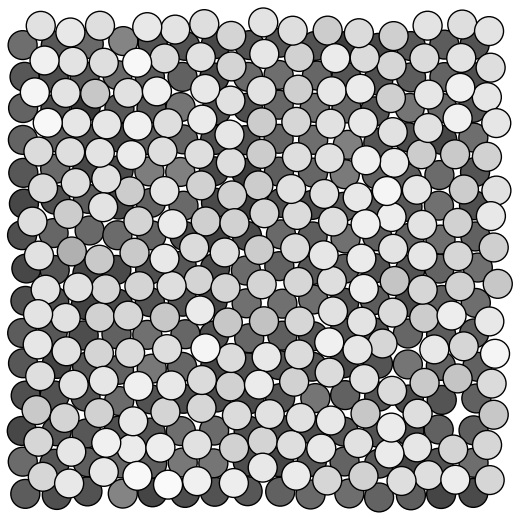
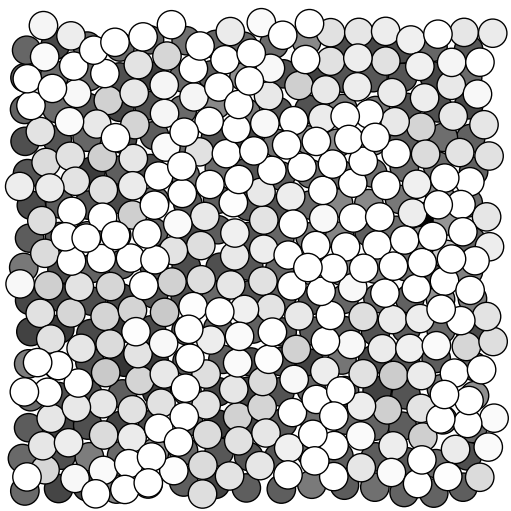
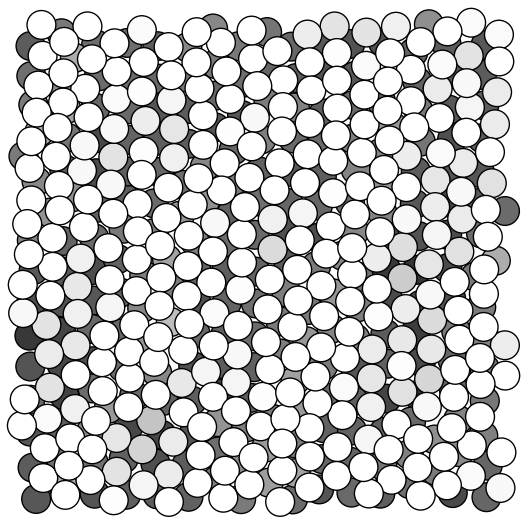
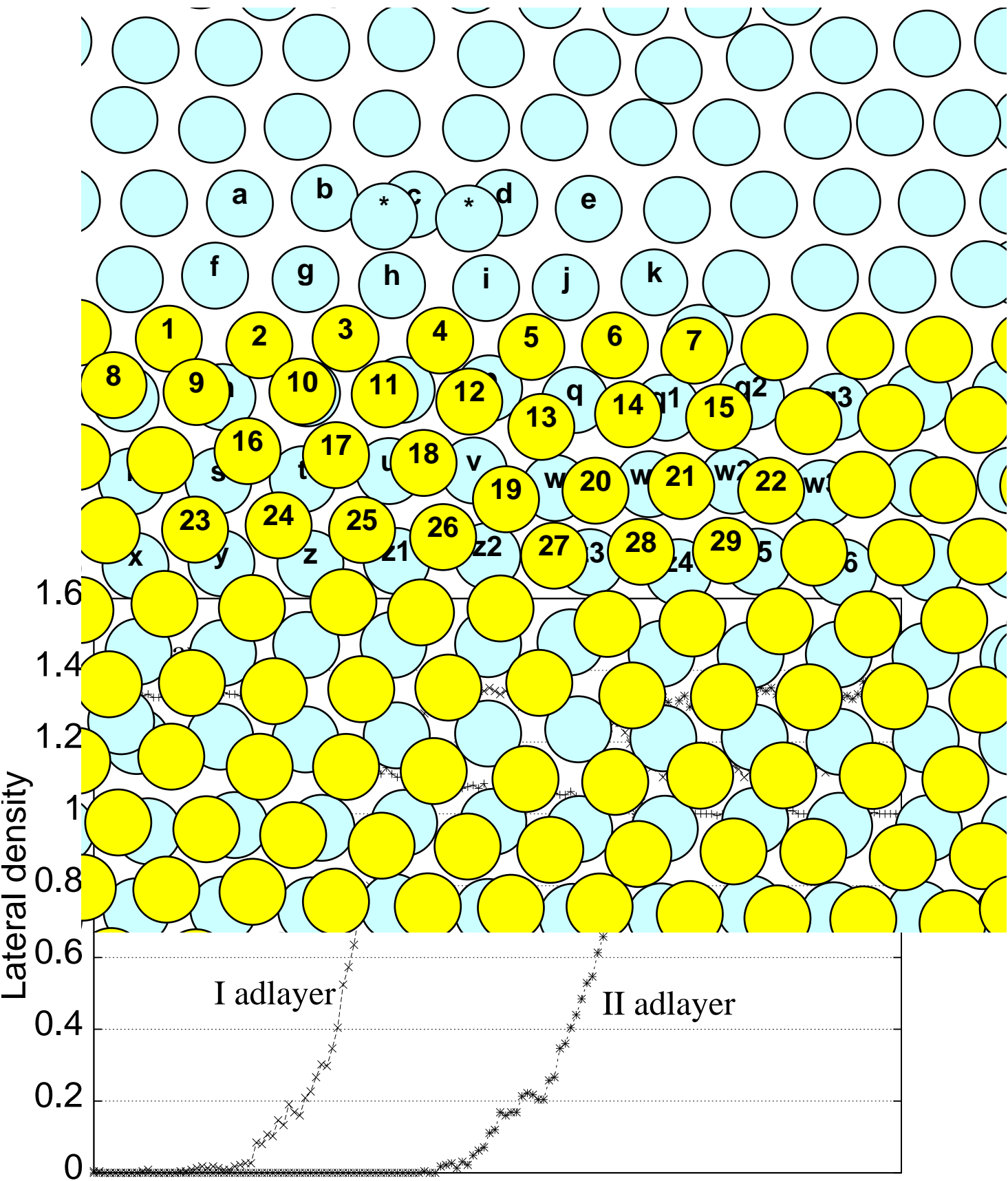
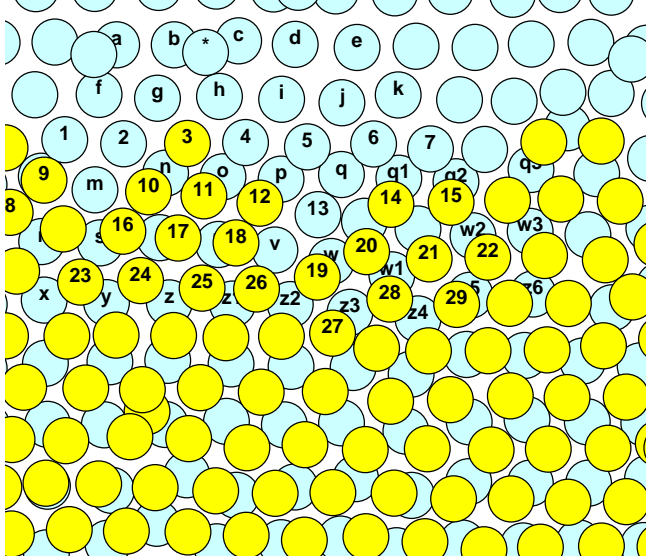
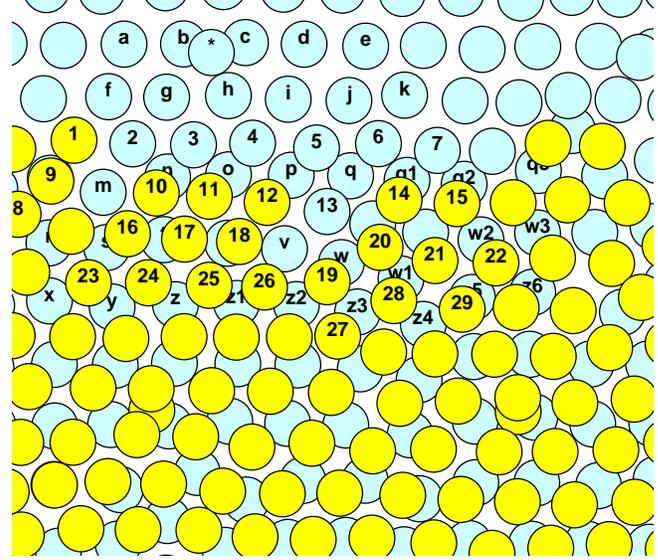
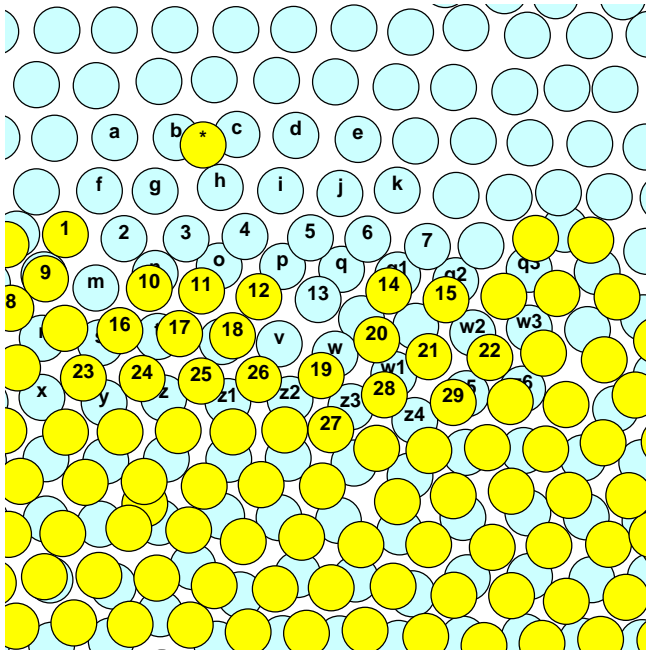
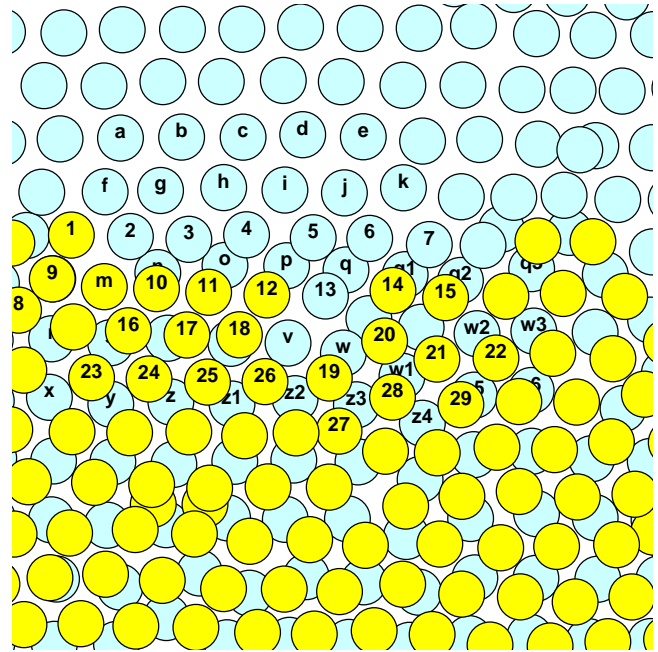
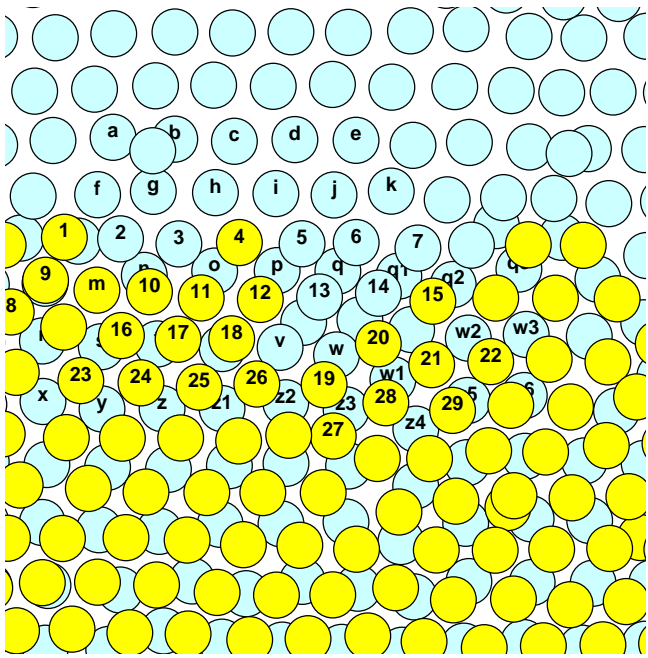
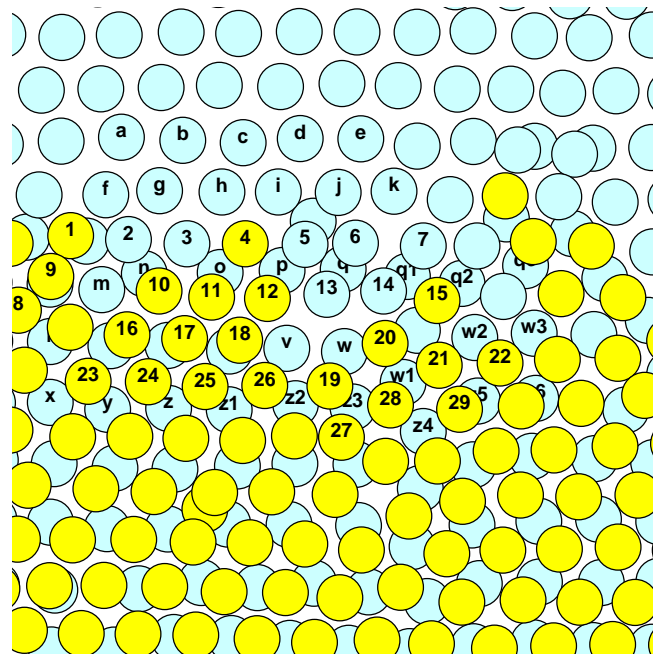
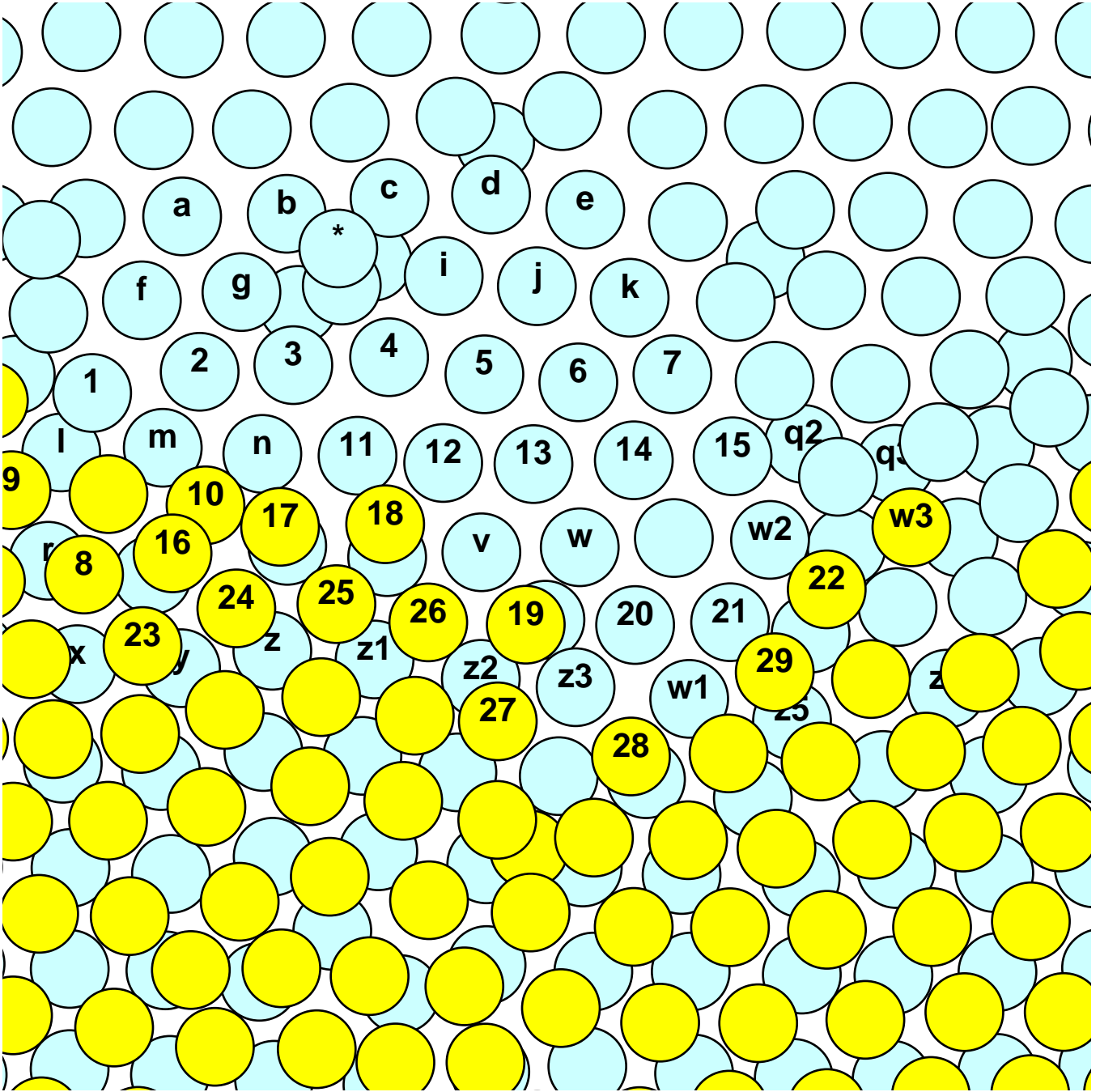
